\documentstyle[axodraw,12pt]{article}

\catcode`@=11
\newcount\@tempcntc
\def\@citex[#1]#2{\if@filesw\immediate\write\@auxout{\string\citation{#2}}\fi
  \@tempcnta\z@\@tempcntb\m@ne\def\@citea{}\@cite{\@for\@citeb:=#2\do
    {\@ifundefined
       {b@\@citeb}{\@citeo\@tempcntb\m@ne\@citea\def\@citea{,}{\bf ?}\@warning
       {Citation `\@citeb' on page \thepage \space undefined}}%
    {\setbox\z@\hbox{\global\@tempcntc0\csname b@\@citeb\endcsname\relax}%
     \ifnum\@tempcntc=\z@ \@citeo\@tempcntb\m@ne
       \@citea\def\@citea{,}\hbox{\csname b@\@citeb\endcsname}%
     \else
      \advance\@tempcntb\@ne
      \ifnum\@tempcntb=\@tempcntc
      \else\advance\@tempcntb\m@ne\@citeo
      \@tempcnta\@tempcntc\@tempcntb\@tempcntc\fi\fi}}\@citeo}{#1}}
\def\@citeo{\ifnum\@tempcnta>\@tempcntb\else\@citea\def\@citea{,}%
  \ifnum\@tempcnta=\@tempcntb\the\@tempcnta\else
   {\advance\@tempcnta\@ne\ifnum\@tempcnta=\@tempcntb \else \def\@citea{--}\fi
    \advance\@tempcnta\m@ne\the\@tempcnta\@citea\the\@tempcntb}\fi\fi}
\catcode`@=12
\begin{document}

\title{\vskip-3cm{\baselineskip14pt
\centerline{\normalsize DESY 99-173\hfill ISSN 0418-9833}
\centerline{\normalsize hep-ph/9911414\hfill}
\centerline{\normalsize November 1999\hfill}}
\vskip1.5cm
Order $\alpha_s^3\ln^2(1/\alpha_s)$ Corrections to Heavy-Quarkonium Creation
and Annihilation} 
\author{{\sc Bernd A. Kniehl} and {\sc Alexander A. Penin}\thanks{Permanent
address: Institute for Nuclear Research, Russian Academy of Sciences,
60th October Anniversary Prospect 7a, Moscow 117312, Russia.}\\
{\normalsize II. Institut f\"ur Theoretische Physik, Universit\"at Hamburg,}\\
{\normalsize Luruper Chaussee 149, 22761 Hamburg, Germany}}

\date{}

\maketitle

\thispagestyle{empty}

\begin{abstract}
In the framework of nonrelativistic QCD, we compute the leading
double-logarith\-mic corrections of order $\alpha_s^3\ln^2(1/\alpha_s)$ to the
heavy-quark-antiquark bound-state wave function at the origin, which
determines the production and annihilation rates of heavy quarkonia.
The phenomenological implications for the top-antitop and $\Upsilon$ systems
are discussed.
\medskip

\end{abstract}

\newpage

\section{Introduction}

Nonrelativistic quantum chromodynamics (NRQCD) \cite{CasLep,LepMag,BodBra} is
a powerful tool for the investigation of heavy-quark threshold dynamics.
Recent developments of the NRQCD effective-theory approach   
\cite{LukMan,Man,Hoa1,GriRot,LukSav,Lab,PinSot1,PinSot2,BenSmi,Gri,CMY}
led to the complete next-to-next-to-leading-order\footnote{In the NRQCD
effective theory, there are two expansion parameters, the strong coupling 
constant $\alpha_s$ and the heavy-quark velocity $\beta$, and the perturbative
order of some correction is determined by their total power.
E.g., terms of $O(\alpha_s^2)$, $O(\alpha_s\beta)$, and $O(\beta^2)$ 
contribute at NNLO.} (NNLO) description of the production of heavy
quark-antiquark pairs at threshold.
Important applications include $\Upsilon$ sum rules and toponium phenomenology
\cite{PY,CzaMel,Beneke,HoaTeu,MelYel,KPP,PP1,Hoa2,BSS,Yak,Nag,PP2,BenSig}.
A review of the recent progress in the perturbative study of heavy
quark-antiquark systems may be found, for example, in Ref.~\cite{rev}.
In view of the surprising significance of the NNLO corrections, it appears
indispensable to also gain control over the next-to-next-to-next-to-leading
order (N$^3$LO) in order to improve the reliability of the theoretical 
predictions and our understanding of the structure and the peculiarities of
the threshold expansion.
 
Some specific classes of N$^3$LO corrections were analyzed in literature.
The one-loop renormalization of the $1/m_q^2$ operators was obtained in
Refs.~\cite{Man,PinSot3}. 
The retardation effects arising from the emission and absorption of virtual
ultrasoft gluons by the heavy quarks were studied in Ref.~\cite{KniPen}.
The leading logarithmic $O\left(\alpha_s^3\ln(1/\alpha_s)\right)$ corrections
to the heavy-quark bound-state energies $E_n$, where $n$ is the principal
quantum number, were obtained in Ref.~\cite{Bra1}.

In this paper, we take the next step in this direction and investigate a 
particular class of N$^3$LO corrections, namely the leading logarithmic
$O\left(\alpha_s^3\ln^2(1/\alpha_s)\right)$ corrections to the wave functions
at the origin $\psi_n(0)$ of the heavy quark-antiquark bound states which are
not generated by the renormalization group (RG).
As is well known, $\psi_n(0)$ are key parameters in the analysis of the
creation and annihilation of heavy quarkonia.
The origin of these logarithmic corrections is the presence of several scales
in the threshold problem.
In fact, $\ln(1/\alpha_s)$ appears as a logarithm of a ratio of scales.
These corrections are related to the anomalous dimensions of the operators in
the effective Hamiltonian.
They can be found by analyzing the divergences of the effective theory or by
direct inspection of the regions of the logarithmic integration.
We shall verify that both methods lead to the same result.
As a by-product of our analysis, we shall also reproduce the
$O\left(\alpha_s^3\ln(1/\alpha_s)\right)$ corrections to $E_n$ recently
obtained in Ref.~\cite{Bra1}.

In N$^3$LO, there are in general also logarithmic corrections of the form
$\alpha_s^3\ln^m(\mu/\alpha_sm_q)$ ($m=1,2,3$), where $\alpha_sm_q$ represents
the soft or potential scales (see Section~2).
These corrections may be directly extracted from the NNLO result via the RG
equation, and they may be resummed by an appropriate choice of the
normalization point, $\mu\approx\alpha_sm_q$.
We note in passing that, in the $\overline{\rm MS}$ scheme, the optimal choice
in NNLO is $\mu\approx\mbox{few units}\times\alpha_sm_q$ \cite{PP1,PP2}.

This paper is organized as follows.
In Section~2, we recall the potential-NRQCD (pNRQCD) formalism, from which
$E_n$ and $\psi_n(0)$ can be extracted.
In Section~3, we evaluate the non-RG
$O\left(\alpha_s^3\ln^2(1/\alpha_s)\right)$ and
$O\left(\alpha_s^3\ln(1/\alpha_s)\right)$ corrections to $\psi_n(0)$ and
$E_n$, respectively, using the effective-theory approach.
In Section~4, we repeat this calculation in the conventional approach, by 
inspecting the logarithmically divergent phase-space integrals.
In Section~5, we present a numerical analysis and discuss phenomenological
implications of our results.
Section~6 contains our conclusions.

\section{Vacuum-polarization function near threshold in\break NRQCD}

We investigate the near-threshold behavior of the vacuum-polarization
function $\Pi(q^2)$ of a heavy-quark vector current
$j_\mu=\bar q\gamma_\mu q$,
\begin{equation}
\left(q_\mu q_\nu-g_{\mu \nu}q^2\right)\Pi(q^2)
=i\int d^4xe^{iq\cdot x}\langle 0|Tj_{\mu}(x)j_{\nu}(0)|0\rangle.
\end{equation}
Its imaginary part is related to the normalized cross section of $q\bar q$
production in $e^+e^-$ annihilation at energy $s=q^2$, 
\begin{equation}
R(s)={\sigma(e^+e^-\rightarrow q\bar q)\over
\sigma(e^+e^-\rightarrow\mu^+\mu^-)},
\end{equation}
by
\begin{equation}
R(s)={12\pi Q_q^2}{\rm Im}\Pi(s+i\epsilon),
\end{equation}
where $Q_q$ is the fractional charge of quark $q$.

Near threshold, the heavy quarks are nonrelativistic, so that one may consider
the quark velocity $\beta=\sqrt{1-4m_q^2/s}$ as a small parameter.
An expansion in $\beta$ may be performed directly in the Lagrangian of QCD by
using the framework of effective field theory.
In the threshold problem, there are four different scales \cite{BenSmi}:
(i) the hard scale (energy and momentum scale like $m_q$);
(ii) the soft scale (energy and momentum scale like $\beta m_q$);
(iii) the potential scale (energy scales like $\beta^2m_q$, while 
momentum scales like $\beta m_q$); and
(iv) the ultrasoft scale (energy and momentum scale like $\beta^2m_q$).
The ultrasoft scale is only relevant for gluons.  
By integrating out the hard scale of QCD, one arrives at the effective theory
of NRQCD \cite{CasLep,LepMag,BodBra}. 
If one also integrates out the soft scale and the potential gluons, one 
obtains the effective theory of pNRQCD, which contains potential quarks and
ultrasoft gluons as active particles \cite{PinSot1}.
The dynamics of the quarks is governed by the effective, nonrelativistic 
Schr\"odinger equation and by their interaction with the ultrasoft gluons.
To get a regular perturbative expansion within pNRQCD, this interaction should
be expanded in multipoles.
The corrections from harder scales are contained in the Wilson coefficients,
leading to an expansion in $\alpha_s$, as well as in the higher-dimensional
operators of the nonrelativistic Hamiltonian, corresponding to an expansion in
$1/m_q$ or $\beta$.

The nonrelativistic expansion in $\alpha_s$ and $\beta$ provides us with the
following representation of the heavy-quark vacuum-polarization function near
threshold:
\begin{equation}
\Pi(E)={N_c\over2m_q^2}C_h(\alpha_s)G(0,0,E)+\ldots,
\end{equation}
where $E=\sqrt{s}-2m_q$ is the $q\bar q$ energy counted from the threshold,
$C_h(\alpha_s)$ is the square of the hard matching coefficient of the 
nonrelativistic vector current, and the ellipsis stands for the higher-order
terms in $\beta$.
$G({\bf x},{\bf y},E)$ is the nonrelativistic Green function, which sums up
the $(\alpha_s/\beta)^n$ terms singular near the threshold.
It is determined by the Schr\"odinger equation which describes the the
propagation of the nonrelativistic quark-antiquark pair in pNRQCD,
\begin{equation}
\left({\cal H}-E\right)G({\bf x},{\bf y},E)
=\delta^{(3)}({\bf x}-{\bf y}),
\end{equation}
where ${\cal H}$ is the nonrelativistic Hamiltonian defined by
\begin{equation}
{\cal H}=-{{\bf \partial}_{\bf x}^2\over m_q}+V(x)+\ldots,\qquad
V(x)=V_C(x)+\ldots\,.
\end{equation}
Here, $V_C(x)=-C_F\alpha_s /x$ is the Coulomb potential, $C_F=4/3$ is the
eigenvalue of the quadratic Casimir operator of the fundamental representation
of the colour group, $x=|{\bf x}|$, and the ellipses stand for the
higher-order terms in $\alpha_s$ and $1/m_q$.
The Green function has the spectral representation
\begin{equation}
G({\bf x},{\bf y},E)
=\sum_{n=1}^\infty{\psi^*_n({\bf x})\psi_n({\bf y})\over E_n-E}
+\int_0^\infty{d^3k\over(2\pi)^3}\,
{\psi^{*}_{\bf k}({\bf x})\psi_{\bf k}({\bf y})\over k^2/m_q-E},
\label{spectr}
\end{equation}
where $\psi_m$ and $\psi_{\bf k}$ are the wave functions of the $q\bar q$
bound and continuum states, respectively.

Below the threshold, the (perturbative) vacuum-polarization function of a
stable heavy quark is essentially determined by the bound-state parameters. 
For the leading-order Coulomb (C) Green function, the energy levels and wave
functions at the origin read
\begin{equation}
E^C_n=-{\lambda_s^2\over m_qn^2},\qquad
\left|\psi_n^C(0)\right|^2={\lambda_s^3\over \pi n^3},
\end{equation}
where $\lambda_s=\alpha_sC_Fm_q/2$.
We are interested in the corrections to $\left|\psi_n^C(0)\right|^2$.
Note that, for the study of bound-state parameters, we have
$\beta\approx\alpha_s$, so that we are only dealing with one expansion
parameter.

\section{Non-RG leading logarithmic corrections from the effective-theory
approach}

The NNLO non-RG leading logarithmic corrections to the wave functions at the
origin are generated by the following operators in the effective
nonrelativistic Hamiltonian:
\begin{equation}
\Delta'{\cal H}=-{C_FC_A\alpha_s^2\over2m_qx^2}
+{C_F\alpha_s\over2m_q^2}\left\{{\bf\partial}_{\bf x}^2,{1\over x}\right\}
+\left(1+{4\over3}{\bf S}^2\right){\pi C_F\alpha_s\over m_q^2}\delta({\bf x})
-{{\bf\partial}_{\bf x}^4\over4m_q^3},
\label{hp}
\end{equation}
where $C_A=3$ is the eigenvalue of the quadratic Casimir operator of the
adjoint representation of the colour group, $\bf S$ represents the spin of the
quark-antiquark system, and $\{.,.\}$ denotes the anticommutator.
The first term of Eq.~(\ref{hp}) is the non-Abelian potential
\cite{GupRad,GRR,TY}, and the rest is the standard Breit potential and the 
kinetic-energy correction.
Note that the representation~(\ref{hp}) of the nonrelativistic Hamiltonian is
not unique and can be related to other representations found in literature by
the use of the equations of motion.
However, the corresponding corrections to the Green function
\cite{HoaTeu,MelYel,Hoa2} are independent of the specific representation. 
In the vicinity of the $n$th bound-state pole, the resulting correction to the
Green function reads \cite{Hoa2}
\begin{equation}
\left.\Delta G(0,0,E)\right|_{E\rightarrow E_n^C}
={\left|\psi_n^C(0)\right|^2\over E_n^C-E}\,{2\pi\alpha_s\over m_q^2}
\left\{\left[4-{4\over 3}S(S+1)\right]C_F+2C_A\right\}G_C(0,0,E)+\ldots,
\label{gfcorr}
\end{equation}
where $G_C({\bf x},{\bf y},E)$ is the Coulomb Green function.
The spin-independent term proportional to $C_F$ in Eq.~(\ref{gfcorr}) comes 
about as $(4-1+1)C_F$, where the contributions arise from the anticommutator,
$\delta$-function, and $1/m_q^3$ terms of Eq.~(\ref{hp}).
Writing
\begin{equation}
|\psi_n(0)|^2=\left|\psi_n^C(0)\right|^2\left(1+\Delta\psi_n^2\right),
\end{equation}
we have
\begin{equation}
\Delta\psi_n^2(0)={2\pi\alpha_s\over m_q^2}
\left\{\left[4-{4\over 3}S(S+1)\right]C_F+2C_A\right\}
G_C(0,0,E_n)+\ldots\,.
\label{wfcorr}
\end{equation}
These corrections are singular, since the Coulomb Green function at the origin
is ultraviolet (UV) divergent.
In dimensional regularization, with $d=4-2\epsilon$ space-time dimensions, 
it takes the form \cite{PP2}
\begin{equation}
G_C(0,0,k)={1\over2\epsilon}\left({\mu\over k}\right)^{2\epsilon}
{m_q\lambda_s\over 2\pi}+\ldots
={m_q\lambda_s\over 2\pi}\left({1\over 2\epsilon}+\ln{\mu\over k}\right)
+\ldots,
\label{gf}
\end{equation}
where $k^2=-m_qE$ and the ellipsis stands for the nonlogarithmic contribution.
The pole term in Eq.~(\ref{gf}) is canceled by the $O\left(\alpha_s^2\right)$
infrared (IR) pole of the hard matching coefficient $C_h(\alpha_s)$
\cite{CzaMel,Beneke}.
Thus, the scale $\mu$ in the logarithm is to be identified with $m_q$, and the
sought correction reads \cite{MelYel,PP1}
\begin{equation}
\Delta'\psi_n^2(0)={C_F\alpha_s^2}
\left\{\left[2-{2\over 3}S(S+1)\right]C_F+C_A\right\}\ln{1\over\alpha_s}.
\label{wfcorrp}
\end{equation}
The residual operators in the NNLO effective nonrelativistic Hamiltonian which
are not contained in Eq.~(\ref{hp}) correspond to the purely perturbative 
corrections to the static Coulomb potential.
The corresponding corrections to the wave functions at the origin
\cite{MelYel,PP1,PP2} contain RG logarithms of the form
$\alpha_s^2\ln^m(\mu/\alpha_sm_q)$ ($m=1,2$), which vanish for
$\mu =\alpha_sm_q$.
This also holds in N$^3$LO for the RG logarithms of the form
$\alpha_s^3\ln^m(\mu/\alpha_sm_q)$ ($m=1,2,3$) because the ultrasoft effects
enter the stage only in N$^3$LO, so that the corresponding running of the
strong coupling constant at the ultrasoft scale only becomes relevant in
N$^4$LO. 
An important point here is that, starting from NNLO, the hard matching
coefficient $C_h(\alpha_s)$ receives a nonvanishing anomalous dimension.
Therefore, starting from N$^3$LO, not only the running of $\alpha_s$ should be
taken into account, but also the effective-theory RG should be used for the
evolution of the hard matching coefficient from $\mu=m_q$ down to
$\mu=\alpha_sm_q$ \cite{BSS}.

At N$^3$LO, the non-RG leading logarithmic corrections to the wave functions
at the origin are produced by the one-loop renormalization of the operators in
Eq.~(\ref{hp}).
In dimensional regularization, the pole part of the correction is
\begin{eqnarray}
\Delta''{\cal H}&=&{1\over 2\epsilon}\,{C_F\alpha_s\over\pi}
\left\{-\left({4\over 3}C_F+{2\over 3}C_A\right){C_A\alpha_s^2\over m_qx^2}
+{2\over 3}\,{C_A\alpha_s\over m_q^2}
\left\{{\bf\partial}^2_{\bf x},{1\over x}\right\}
-\left({16\over3}C_F-{8\over3}C_A\right){\pi\alpha_s\over m_q^2}
\delta({\bf x})
\right.\nonumber\\
&&{}+\left.
\left[{2\over3}C_F+\left({17\over3}-{7\over3}{\bf S}^2\right)C_A\right]
{\pi\alpha_s\over m_q^2}\delta({\bf x})\right\},
\label{hpp}
\end{eqnarray}
where the first three terms contained within the parentheses represent the IR
divergence, while the fourth one embodies the UV divergence of the potential.
The IR poles are canceled by the ultrasoft contribution, with the
characteristic scale $\alpha_s^2m_q$, and may be read off from
Refs.~\cite{KniPen,Bra1}, while the UV poles are canceled by the IR poles of
the hard coefficients and may be extracted from Refs.~\cite{Man,PinSot3}.  
Evaluating the corrections to the wave functions at the origin due to
Eq.~(\ref{hpp}) in the same way as Eq.~(\ref{wfcorr}) was obtained, we find
the N$^3$LO non-RG leading logarithmic corrections to the wave functions at
the origin to be
\begin{equation}
\Delta''\psi_n^2(0)=-{C_F\alpha_s^3\over\pi}
\left\{{3\over2}C_F^2+\left[{41\over12}-{7\over12}S(S+1)\right]C_FC_A
+{2\over 3}C_A^2\right\}\ln^2{1\over\alpha_s}.
\label{wfcorrpp}
\end{equation}
In the derivation of this result, the factor $(\mu/k)^{2\epsilon}$ in
Eq.~(\ref{gf}) needed to be expanded up to $O(\epsilon^2)$.
The contributions to Eq.~(\ref{wfcorrpp}) related to the IR and UV poles of
Eq.~(\ref{hpp}) are of opposite signs.
This may be understood by observing that the IR poles introduce a factor
$\ln(E_0^C/\lambda_s)\approx\ln\alpha_s$, while the UV poles contribute a
factor $\ln(m_q/\lambda_s)\approx\ln(1/\alpha_s)$. 
The Abelian $C_F^3$ term in Eq.~(\ref{wfcorrpp}) agrees with the QED result of
Ref.~\cite{Kar}.
In contrast to the QED case, the leading logarithmic QCD corrections are spin
dependent.

In the remainder of this section, we revisit the leading logarithmic 
corrections to the bound-state energy levels, which may be obtained in a way
similar to the case of the wave functions at the origin.
These corrections start from N$^3$LO, since there are no relevant
singularities in NNLO.
In addition to Eq.~(\ref{hpp}), we now have to take into account the
IR-singular contribution to the Coulomb potential \cite{ADM,Bra2},
\begin{equation}
-{1\over 2\epsilon}\,{C_FC_A^3\alpha_s^4\over 12x},
\end{equation}
which is canceled by the ultrasoft contribution \cite{KniPen}.
Obviously, this term gives no contributions to the wave functions. 
Writing
\begin{equation}
E_n=E^C_n+\Delta E_n,
\end{equation}
we find
\begin{eqnarray}
\Delta E_n&=&-E_n{\alpha_s^3\over\pi}\left\{{3\over n}C_F^3
+\left[{41\over6n}-{7\over6n}S(S+1)-{2\over3n^2}\right]C_F^2C_A
+{4\over3n}C_FC_A^2+{1\over6}C_A^3\right\}
\nonumber\\
&&{}\times\ln{1\over\alpha_s},
\label{encorrpp}
\end{eqnarray}
in agreement with the result for $l=0$ of Ref.~\cite{Bra1}.

\section{Non-RG leading logarithmic corrections from the traditional
approach}

An alternative method of finding the leading logarithmic corrections is to
directly inspect the regions of logarithmic integration
\cite{Kar,Caswel,EMK,Melnik}.
Let us first consider the NNLO corrections to the wave functions.
Substituting the continuum part of the spectral representation~(\ref{spectr})
into Eq.~(\ref{wfcorr}), we obtain
\begin{equation}
\Delta'\psi_n^2(0)={2\pi\alpha_s\over m_q^2}
\left\{\left[4-{4\over 3}S(S+1)\right]C_F+2C_A\right\}
\int_0^\infty{d^3k\over(2\pi)^3}\,
{\left|\psi_{\bf k}^C(0)\right|^2\over k^2/m_q-E_n}.
\label{nnlow}
\end{equation}
The integral over $k$ logarithmically diverges at large momentum and should be
cut at scale $m_q$, where the nonrelativistic approximation becomes
inapplicable.
At low momentum, the logarithmic integration over $k$ is effectively cut at
the Coulomb scale $\alpha_sm_q$.
Inserting in Eq.~(\ref{nnlow}) the well-known expression for the Coulomb wave
function at the origin,
\begin{equation}
\left|\psi_{\bf k}^C(0)\right|^2={2\pi\lambda_s\over k}\,
{1\over 1-\exp{(-2\pi\lambda_s/k)}}
={1}+{\pi\lambda_s\over k}+\ldots,
\end{equation}
where only the second term is relevant for our purposes,
we find, with logarithmic accuracy,
\begin{equation}
\Delta'\psi_n^2(0)={C_F\alpha_s^2}
\left\{\left[2-{2\over 3}S(S+1)\right]C_F+C_A\right\}
\int_{\alpha_sm_q}^{m_q}{dk\over k}+\ldots\,.
\label{wfintp}
\end{equation}
Performing in Eq.~(\ref{wfintp}) the integration over $k$, we recover
Eq.~(\ref{wfcorrp}).

The N$^3$LO leading (double) logarithmic corrections to the wave function at
the origin are of the form
\begin{eqnarray}
\Delta''\psi_n^2(0)&=&-{C_F\alpha_s^2}
\left\{\left({8\over 3}C_F^2+4C_AC_F+{4\over 3}C_A^2\right)
\int_{\alpha_sm_q}^{m_q}{dk\over k}
\int_{-E}^{k}{dk'\over k'}
\right.\nonumber\\
&&{}+\left.
\left[{1\over 3}C_F^2+\left({17\over 6}-{7\over 6}S(S+1)\right)C_FC_A\right]
\int_{\alpha_sm_q}^{m_q}{dk\over k}\int_{k}^{m_q}{dk'\over k'}\right\}.
\label{wfintpp}
\end{eqnarray}
Here $k$ is the momentum of the potential heavy quark, which can be as small
as $\alpha_sm_q$, while $k'$ is the momentum of the virtual gluon, which can
even be ultrasoft, of order $\alpha_s^2m_q$.
The integrals over $k$ have the same origin as those in Eq.~(\ref{wfintp}).
The integrals over $k'$ represent the logarithmic corrections to the potential
due to the virtual-gluon exchanges \cite{Man,PinSot3,KniPen}.  
In the first integral on the right-hand side of Eq.~(\ref{wfintpp}), the quark 
momentum $k$ plays the role of an UV cutoff, and the quark energy $-E=k^2/m_q$
acts as an IR cutoff, so that this contribution corresponds to the IR poles of
Eq.~(\ref{hpp}).
In the second integral, the momentum $k$ acts as an IR cutoff, and this
contribution corresponds to the UV poles of Eq.~(\ref{hpp}).
Integrating  Eq.~(\ref{wfintpp}) over $k'$ and $k$, we arrive at
Eq.~(\ref{wfcorrpp}).
A similar analysis for QED bound states may be found in Ref.~\cite{Kar}.

The N$^3$LO leading (single) logarithmic corrections to the energy levels may
be obtained by the method introduced in Refs.~\cite{Caswel,EMK,Melnik}, where
the regions of the virtual-photon momentum which lead to logarithmic
contributions have been studied in the QED case. 
In this way, we obtain
\begin{eqnarray}
\Delta E_n&=&-E_n{\alpha_s^3\over\pi}
\left\{\left[{8\over3n}C_F^3+\left({31\over6n}-{2\over3n^2}\right)C_F^2C_A
+{4\over3n}C_FC_A^2+{1\over6}C_A^3\right]
\int_{\alpha_s^2m_q}^{\alpha_sm_q}{dk'\over k'}\right.
\nonumber\\
&&{}\left.
+\left[{1\over3n}C_F^3
+\left({17\over6n}-{7\over6n}S(S+1)-{2\over3n^2}\right)C_F^2C_A\right]
\int_{\alpha_sm_q}^{m_q}{dk'\over k'}\right\},
\label{enintpp}
\end{eqnarray}
where again the first (second) integral corresponds to the IR (UV) poles of
Eq.~(\ref{hpp}).
After integration, we recover Eq.~(\ref{encorrpp}).

\section{Phenomenological applications}

Our results affect two processes of primary interest, namely the threshold 
production of top and bottom quark-antiquark pairs.
The relatively large width $\Gamma_t$ of the top quark serves as an efficient
IR cutoff for long-distance effects.
Because the relevant scale $\sqrt{m_t\Gamma_t}$ is much larger than the 
asymptotic scale parameter $\Lambda_{\rm QCD}$, but comparable to the Coulomb
scale, the cross section in the threshold region may be described by the
NRQCD perturbation expansion if singular Coulomb effects are properly taken
into account \cite{FK}.
In order to analyze the significance of the N$^3$LO leading logarithmic
corrections to the cross section, we start from the NNLO calculation of
Ref.~\cite{PP2} and add the contributions from Eqs.~(\ref{wfcorrpp}) and
(\ref{encorrpp}).
In Fig.~1, the normalized cross section $R$ thus obtained is compared with the
pure NNLO result.
The input parameters are taken to be $\alpha_s(M_Z)=0.118$, $m_t=175$~GeV, and
$\Gamma_t=1.43$~GeV.
The soft renormalization scale is determined from the condition
$\mu_s=2\alpha_s(\mu_s)m_t$, and hard renormalization scale is chosen to be
$\mu_h=m_t$.
The effect of the N$^3$LO leading logarithms is twofold.
The normalization of the cross section is reduced by about 7\% around the 1S
peak and below, and the energy gap between the 1S peak and the nominal 
threshold is decreased by roughly 10\%.

In the case of bottom quark-antiquark production, the nonperturbative effects
are much more significant, and one is led to use the sum rule approach
\cite{NSVZ,Vol,VolZai} to get them under control.
Specifically, appealing quark-hadron duality, one matches the theoretical 
results for the moments of the spectral density,
\begin{eqnarray}
{\cal M}_n&=&
\left.{12\pi^2\over n!}(4m_q^2)^n{d^n\over ds^n}\Pi(s)\right|_{s=0}
\nonumber\\
&=&(4m_q^2)^n\int_0^\infty ds{R(s)\over s^{n+1}},
\label{spec}
\end{eqnarray}
with their experimental counterparts evaluated from the expressions in the 
second line of Eq.~(\ref{spec}).
For large $n$, the moments are saturated by the near-threshold region. 
Then, the main contribution to the experimental moments comes from the
$\Upsilon$ resonances, which are measured with high precision.
On the other hand, for $n$ of $O\left(1/\alpha_s^2\right)$, the Coulomb
effects should be properly taken into account on the theoretical side. 
In order to analyze the N$^3$LO leading logarithmic corrections to the
$\Upsilon$ sum rules, we upgrade the NNLO result of Ref.~\cite{PP1} by
including Eqs.~(\ref{wfcorrpp}) and (\ref{encorrpp}).
We fix the strong coupling constant by $\alpha_s(M_Z)=0.118$ and focus on the
determination of the bottom-quark mass.
At present, this appears to be the most interesting application of
Eq.~(\ref{spec}).
We find that the inclusion of the N$^3$LO leading logarithms in the sum rules
leads to a reduction of the extracted mass value by approximately 30~MeV for
moderate values of $n$, $5<n<15$, and if the soft renormalization point 
$\mu_s$ is chosen to be $\mu_s=4\alpha_s(\mu_s)m_b$. 
This result does not depend on whether the energy denominators of the Green
function are expanded around the Coulomb values or not (see Ref.~\cite{PP1}
for details) and on which mass parameter, pole or $\overline{\rm MS}$ mass, is
considered.  
On the other hand, the result essentially depends on $\mu_s$ because the
$\mu_s$ dependence of $\alpha_s$ is compensated only by higher-order terms.
For example, for $\mu_s=2\alpha_s(\mu_s)m_b$, the correction to the mass
parameter reaches $-70$~MeV. 
However, the perturbative result can hardly be trusted at such a low
renormalization point \cite{PP1,PP2}.  

For completeness, we also present the numerical corrections to the parameters
of the 1S $\Upsilon$ resonance.
For $m_b=4.8$~GeV and $\mu_s=4\alpha_s(\mu_s)m_b$, the wave-function
correction is $-19$\%, and the correction to the binding energy is 25~MeV.

\section{Discussion and conclusions}

We studied a special class of NNLO and N$^3$LO corrections to the key
parameters of heavy quark-antiquark bound states, namely those which are
enhanced by a maximum power of $\ln(1/\alpha_s)$ and are not generated by the
RG.
By contrast, the RG logarithms are well known and may be resummed by an
appropriate scale choice. 
Such non-RG leading logarithmic corrections first arise for the wave functions
at the origin in NNLO \cite{MelYel,PP1} and for the energy levels in N$^3$LO
\cite{Bra1}.
Specifically, they are of the forms $\alpha_s^2\ln(1/\alpha_s)$ and
$\alpha_s^3\ln(1/\alpha_s)$, respectively.
We confirmed these results and completed the knowledge of non-RG leading
logarithmic corrections in N$^3$LO by computing the
$O\left(\alpha_s^3\ln^2(1/\alpha_s)\right)$ corrections to the wave functions
at the origin.

We applied our result to top and bottom quark-antiquark production at 
threshold.
In the case of top, the resulting correction to the production cross section
near the $1S$ peak reaches 10\%. 
In the case of bottom, the mass extracted from the $\Upsilon$ sum rules is
shifted by approximately $-30$~MeV.
The latter value is comparable to the uncertainty of $\pm60$~MeV, which is
usually assigned to this kind of bottom-quark mass determination on the basis
of the renormalization scale dependence of the strong coupling constant in the
NNLO result.
In fact, the non-RG leading logarithmic contributions are about two times
smaller than the contribution from the N$^3$LO RG logarithms, which may be
estimated from the renormalization scale dependence of the NNLO result
\cite{PP1,PP2}.
Thus, the resummation of the RG logarithms \cite{Nag} seems to be very
justified. 
At the same time, the scale of the corrections is close to the estimate given
in Ref.~\cite{KniPen} on the basis of the analysis of the ultrasoft
contributions.

Although, in contrast to QED, $\ln(1/\alpha_s)$ is not a big number,
especially for the case of bottom, the leading logarithmic terms can be
considered as typical representatives of the N$^3$LO corrections.
For comparison, at NNLO, the leading logarithmic term accounts for
approximately one half (third) of the total correction to the $n=0$ wave
function at the origin in the case of top (bottom).
Obviously, the N$^3$LO corrections are comparable to the NNLO ones and reach
10\% in magnitude, even in the case of top, where $\alpha_s\approx1/10$.
This tells us that the NRQCD threshold expansion is not a fast convergent
series for the physical value of the strong coupling constant.

A final comment refers to the resummation of the non-RG leading logarithmic
corrections to the wave functions at the origin.
A part of the non-RG leading logarithmic corrections to the heavy-quark
threshold cross section was resummed in Ref.~\cite{BSS} by using the RG
equation of the effective theory for the evolution of the hard matching
coefficient $C_h(\alpha_s)$ from scale $m_q$ down to scale $\beta m_q$.
This evolution equation is obtained by studying the dependence of
$C_h(\alpha_s)$ on the scale $\mu$, which cancels the $\mu$ dependence of
Eqs.~(\ref{wfcorr}) and (\ref{gf}).
This effectively sums up the higher-order NRQCD corrections due to the
tree-level operators of Eq.~(\ref{hp}).
The terms resummed in this way are of the form 
$\alpha_s(\alpha_s\ln(1/\beta))^{2n-1}$, with $n=1,2,\ldots$,
{\it i.e.}\ they include only even powers of $\alpha_s$.
For the bound-state parameters, we have $\beta\approx\alpha_s$, and 
Eq.~(\ref{wfcorrpp}) gives the first term of this series.
In order to resum all correction of the form
$\alpha_s(\alpha_s\ln(1/\alpha_s))^n$, it is necessary to add the terms with
odd powers of $\alpha_s$, of the form 
$\alpha_s(\alpha_s\ln(1/\alpha_s))^{2n}$, which are generated by
Eq.~(\ref{hpp}). 
For this end, one has to take into account not only the evolution of
$C_h(\alpha_s)$, but also the evolution of the potential~(\ref{hp}) from the
hard scale down to the ultrasoft scale.
Numerically, however, the effect of the resummation may not be essential for
phenomenological applications. 

\vspace{1cm}
\noindent
{\bf Acknowledgements}
\smallskip

\noindent
This work was supported by the Bundesministerium f\"ur Bildung und Forschung
under Contract No.\ 05~HT9GUA~3, and by the European Commission through the
Research Training Network {\it Quantum Chromodynamics and the Deep Structure
of Elementary Particles} under Contract No.\ ERBFMRXCT980194.
The work of A.A.P. was supported in part by the Volkswagen Foundation under
Contract No.\ I/73611, and by the Russian Academy of Sciences through Grant
No.\ 37.

\newpage

\begin{center}

% GNUPLOT: LaTeX picture
\setlength{\unitlength}{0.240900pt}
\ifx\plotpoint\undefined\newsavebox{\plotpoint}\fi
\begin{picture}(1500,990)(0,0)
\font\gnuplot=cmr10 at 10pt
\gnuplot
\sbox{\plotpoint}{\rule[-0.200pt]{0.400pt}{0.400pt}}%
\put(181.0,163.0){\rule[-0.200pt]{4.818pt}{0.400pt}}
\put(161,163){\makebox(0,0)[r]{0.3}}
\put(1460.0,163.0){\rule[-0.200pt]{4.818pt}{0.400pt}}
\put(181.0,250.0){\rule[-0.200pt]{4.818pt}{0.400pt}}
\put(161,250){\makebox(0,0)[r]{0.4}}
\put(1460.0,250.0){\rule[-0.200pt]{4.818pt}{0.400pt}}
\put(181.0,338.0){\rule[-0.200pt]{4.818pt}{0.400pt}}
\put(161,338){\makebox(0,0)[r]{0.5}}
\put(1460.0,338.0){\rule[-0.200pt]{4.818pt}{0.400pt}}
\put(181.0,425.0){\rule[-0.200pt]{4.818pt}{0.400pt}}
\put(161,425){\makebox(0,0)[r]{0.6}}
\put(1460.0,425.0){\rule[-0.200pt]{4.818pt}{0.400pt}}
\put(181.0,512.0){\rule[-0.200pt]{4.818pt}{0.400pt}}
\put(161,512){\makebox(0,0)[r]{0.7}}
\put(1460.0,512.0){\rule[-0.200pt]{4.818pt}{0.400pt}}
\put(181.0,600.0){\rule[-0.200pt]{4.818pt}{0.400pt}}
\put(161,600){\makebox(0,0)[r]{0.8}}
\put(1460.0,600.0){\rule[-0.200pt]{4.818pt}{0.400pt}}
\put(181.0,687.0){\rule[-0.200pt]{4.818pt}{0.400pt}}
\put(161,687){\makebox(0,0)[r]{0.9}}
\put(1460.0,687.0){\rule[-0.200pt]{4.818pt}{0.400pt}}
\put(181.0,774.0){\rule[-0.200pt]{4.818pt}{0.400pt}}
\put(161,774){\makebox(0,0)[r]{1}}
\put(1460.0,774.0){\rule[-0.200pt]{4.818pt}{0.400pt}}
\put(181.0,862.0){\rule[-0.200pt]{4.818pt}{0.400pt}}
\put(161,862){\makebox(0,0)[r]{1.1}}
\put(1460.0,862.0){\rule[-0.200pt]{4.818pt}{0.400pt}}
\put(181.0,949.0){\rule[-0.200pt]{4.818pt}{0.400pt}}
\put(161,949){\makebox(0,0)[r]{1.2}}
\put(1460.0,949.0){\rule[-0.200pt]{4.818pt}{0.400pt}}
\put(181.0,163.0){\rule[-0.200pt]{0.400pt}{4.818pt}}
\put(181,122){\makebox(0,0){-5}}
\put(181.0,929.0){\rule[-0.200pt]{0.400pt}{4.818pt}}
\put(441.0,163.0){\rule[-0.200pt]{0.400pt}{4.818pt}}
\put(441,122){\makebox(0,0){-4}}
\put(441.0,929.0){\rule[-0.200pt]{0.400pt}{4.818pt}}
\put(701.0,163.0){\rule[-0.200pt]{0.400pt}{4.818pt}}
\put(701,122){\makebox(0,0){-3}}
\put(701.0,929.0){\rule[-0.200pt]{0.400pt}{4.818pt}}
\put(960.0,163.0){\rule[-0.200pt]{0.400pt}{4.818pt}}
\put(960,122){\makebox(0,0){-2}}
\put(960.0,929.0){\rule[-0.200pt]{0.400pt}{4.818pt}}
\put(1220.0,163.0){\rule[-0.200pt]{0.400pt}{4.818pt}}
\put(1220,122){\makebox(0,0){-1}}
\put(1220.0,929.0){\rule[-0.200pt]{0.400pt}{4.818pt}}
\put(1480.0,163.0){\rule[-0.200pt]{0.400pt}{4.818pt}}
\put(1480,122){\makebox(0,0){0}}
\put(1480.0,929.0){\rule[-0.200pt]{0.400pt}{4.818pt}}
\put(181.0,163.0){\rule[-0.200pt]{312.929pt}{0.400pt}}
\put(1480.0,163.0){\rule[-0.200pt]{0.400pt}{189.347pt}}
\put(181.0,949.0){\rule[-0.200pt]{312.929pt}{0.400pt}}
\put(1,556){\makebox(0,0){$R(E)$}}
\put(830,1){\makebox(0,0){$E$ (GeV)}}
\put(181.0,163.0){\rule[-0.200pt]{0.400pt}{189.347pt}}
\sbox{\plotpoint}{\rule[-0.500pt]{1.000pt}{1.000pt}}%
\put(181,185){\usebox{\plotpoint}}
\put(181.00,185.00){\usebox{\plotpoint}}
\put(199.27,194.84){\usebox{\plotpoint}}
\put(217.55,204.68){\usebox{\plotpoint}}
\put(235.73,214.68){\usebox{\plotpoint}}
\put(253.41,225.56){\usebox{\plotpoint}}
\put(271.08,236.44){\usebox{\plotpoint}}
\put(288.18,248.20){\usebox{\plotpoint}}
\put(304.99,260.37){\usebox{\plotpoint}}
\put(321.83,272.50){\usebox{\plotpoint}}
\put(338.31,285.10){\usebox{\plotpoint}}
\put(354.15,298.51){\usebox{\plotpoint}}
\put(369.99,311.92){\usebox{\plotpoint}}
\put(385.47,325.74){\usebox{\plotpoint}}
\put(400.72,339.82){\usebox{\plotpoint}}
\put(415.45,354.45){\usebox{\plotpoint}}
\put(430.04,369.20){\usebox{\plotpoint}}
\put(444.17,384.41){\usebox{\plotpoint}}
\put(458.29,399.62){\usebox{\plotpoint}}
\put(472.21,415.01){\usebox{\plotpoint}}
\put(485.59,430.88){\usebox{\plotpoint}}
\put(498.68,446.98){\usebox{\plotpoint}}
\put(511.76,463.09){\usebox{\plotpoint}}
\put(524.64,479.37){\usebox{\plotpoint}}
\put(537.24,495.86){\usebox{\plotpoint}}
\put(549.68,512.48){\usebox{\plotpoint}}
\put(561.97,529.20){\usebox{\plotpoint}}
\put(574.45,545.78){\usebox{\plotpoint}}
\put(586.60,562.61){\usebox{\plotpoint}}
\put(598.69,579.48){\usebox{\plotpoint}}
\multiput(610,596)(12.152,16.826){2}{\usebox{\plotpoint}}
\put(634.73,630.25){\usebox{\plotpoint}}
\put(646.89,647.07){\usebox{\plotpoint}}
\put(659.04,663.90){\usebox{\plotpoint}}
\put(671.53,680.47){\usebox{\plotpoint}}
\put(684.49,696.68){\usebox{\plotpoint}}
\put(697.58,712.79){\usebox{\plotpoint}}
\put(711.04,728.59){\usebox{\plotpoint}}
\put(724.63,744.27){\usebox{\plotpoint}}
\put(739.12,759.12){\usebox{\plotpoint}}
\put(754.44,773.11){\usebox{\plotpoint}}
\put(771.08,785.52){\usebox{\plotpoint}}
\put(788.47,796.83){\usebox{\plotpoint}}
\put(806.72,806.66){\usebox{\plotpoint}}
\put(826.45,812.95){\usebox{\plotpoint}}
\put(846.92,816.00){\usebox{\plotpoint}}
\put(867.68,816.00){\usebox{\plotpoint}}
\put(888.00,812.15){\usebox{\plotpoint}}
\put(907.84,806.05){\usebox{\plotpoint}}
\put(927.05,798.21){\usebox{\plotpoint}}
\put(945.89,789.51){\usebox{\plotpoint}}
\put(964.06,779.50){\usebox{\plotpoint}}
\put(982.03,769.14){\usebox{\plotpoint}}
\put(999.84,758.48){\usebox{\plotpoint}}
\put(1017.70,747.93){\usebox{\plotpoint}}
\put(1035.62,737.46){\usebox{\plotpoint}}
\put(1053.81,727.48){\usebox{\plotpoint}}
\put(1072.34,718.15){\usebox{\plotpoint}}
\put(1090.78,708.64){\usebox{\plotpoint}}
\put(1109.81,700.38){\usebox{\plotpoint}}
\put(1129.19,692.94){\usebox{\plotpoint}}
\put(1148.86,686.36){\usebox{\plotpoint}}
\put(1168.80,680.75){\usebox{\plotpoint}}
\put(1188.90,675.79){\usebox{\plotpoint}}
\put(1209.22,671.66){\usebox{\plotpoint}}
\put(1229.74,668.50){\usebox{\plotpoint}}
\put(1250.37,666.33){\usebox{\plotpoint}}
\put(1270.98,664.08){\usebox{\plotpoint}}
\put(1291.70,663.00){\usebox{\plotpoint}}
\put(1312.41,662.00){\usebox{\plotpoint}}
\put(1333.17,662.00){\usebox{\plotpoint}}
\put(1353.93,662.00){\usebox{\plotpoint}}
\put(1374.68,662.00){\usebox{\plotpoint}}
\put(1395.40,663.00){\usebox{\plotpoint}}
\put(1416.15,663.09){\usebox{\plotpoint}}
\put(1436.87,664.00){\usebox{\plotpoint}}
\put(1457.58,665.28){\usebox{\plotpoint}}
\put(1478.31,666.00){\usebox{\plotpoint}}
\put(1480,666){\usebox{\plotpoint}}
\sbox{\plotpoint}{\rule[-0.400pt]{0.800pt}{0.800pt}}%
\put(181,165){\usebox{\plotpoint}}
\multiput(181.00,166.39)(1.244,0.536){5}{\rule{1.933pt}{0.129pt}}
\multiput(181.00,163.34)(8.987,6.000){2}{\rule{0.967pt}{0.800pt}}
\multiput(194.00,172.39)(1.244,0.536){5}{\rule{1.933pt}{0.129pt}}
\multiput(194.00,169.34)(8.987,6.000){2}{\rule{0.967pt}{0.800pt}}
\multiput(207.00,178.39)(1.244,0.536){5}{\rule{1.933pt}{0.129pt}}
\multiput(207.00,175.34)(8.987,6.000){2}{\rule{0.967pt}{0.800pt}}
\multiput(220.00,184.40)(1.000,0.526){7}{\rule{1.686pt}{0.127pt}}
\multiput(220.00,181.34)(9.501,7.000){2}{\rule{0.843pt}{0.800pt}}
\multiput(233.00,191.40)(1.000,0.526){7}{\rule{1.686pt}{0.127pt}}
\multiput(233.00,188.34)(9.501,7.000){2}{\rule{0.843pt}{0.800pt}}
\multiput(246.00,198.40)(1.000,0.526){7}{\rule{1.686pt}{0.127pt}}
\multiput(246.00,195.34)(9.501,7.000){2}{\rule{0.843pt}{0.800pt}}
\multiput(259.00,205.40)(1.000,0.526){7}{\rule{1.686pt}{0.127pt}}
\multiput(259.00,202.34)(9.501,7.000){2}{\rule{0.843pt}{0.800pt}}
\multiput(272.00,212.40)(0.847,0.520){9}{\rule{1.500pt}{0.125pt}}
\multiput(272.00,209.34)(9.887,8.000){2}{\rule{0.750pt}{0.800pt}}
\multiput(285.00,220.40)(0.847,0.520){9}{\rule{1.500pt}{0.125pt}}
\multiput(285.00,217.34)(9.887,8.000){2}{\rule{0.750pt}{0.800pt}}
\multiput(298.00,228.40)(0.847,0.520){9}{\rule{1.500pt}{0.125pt}}
\multiput(298.00,225.34)(9.887,8.000){2}{\rule{0.750pt}{0.800pt}}
\multiput(311.00,236.40)(0.737,0.516){11}{\rule{1.356pt}{0.124pt}}
\multiput(311.00,233.34)(10.186,9.000){2}{\rule{0.678pt}{0.800pt}}
\multiput(324.00,245.40)(0.737,0.516){11}{\rule{1.356pt}{0.124pt}}
\multiput(324.00,242.34)(10.186,9.000){2}{\rule{0.678pt}{0.800pt}}
\multiput(337.00,254.40)(0.737,0.516){11}{\rule{1.356pt}{0.124pt}}
\multiput(337.00,251.34)(10.186,9.000){2}{\rule{0.678pt}{0.800pt}}
\multiput(350.00,263.40)(0.654,0.514){13}{\rule{1.240pt}{0.124pt}}
\multiput(350.00,260.34)(10.426,10.000){2}{\rule{0.620pt}{0.800pt}}
\multiput(363.00,273.40)(0.654,0.514){13}{\rule{1.240pt}{0.124pt}}
\multiput(363.00,270.34)(10.426,10.000){2}{\rule{0.620pt}{0.800pt}}
\multiput(376.00,283.40)(0.654,0.514){13}{\rule{1.240pt}{0.124pt}}
\multiput(376.00,280.34)(10.426,10.000){2}{\rule{0.620pt}{0.800pt}}
\multiput(389.00,293.40)(0.589,0.512){15}{\rule{1.145pt}{0.123pt}}
\multiput(389.00,290.34)(10.623,11.000){2}{\rule{0.573pt}{0.800pt}}
\multiput(402.00,304.40)(0.589,0.512){15}{\rule{1.145pt}{0.123pt}}
\multiput(402.00,301.34)(10.623,11.000){2}{\rule{0.573pt}{0.800pt}}
\multiput(415.00,315.41)(0.536,0.511){17}{\rule{1.067pt}{0.123pt}}
\multiput(415.00,312.34)(10.786,12.000){2}{\rule{0.533pt}{0.800pt}}
\multiput(428.00,327.41)(0.536,0.511){17}{\rule{1.067pt}{0.123pt}}
\multiput(428.00,324.34)(10.786,12.000){2}{\rule{0.533pt}{0.800pt}}
\multiput(441.00,339.41)(0.492,0.509){19}{\rule{1.000pt}{0.123pt}}
\multiput(441.00,336.34)(10.924,13.000){2}{\rule{0.500pt}{0.800pt}}
\multiput(454.00,352.41)(0.492,0.509){19}{\rule{1.000pt}{0.123pt}}
\multiput(454.00,349.34)(10.924,13.000){2}{\rule{0.500pt}{0.800pt}}
\multiput(467.00,365.41)(0.492,0.509){19}{\rule{1.000pt}{0.123pt}}
\multiput(467.00,362.34)(10.924,13.000){2}{\rule{0.500pt}{0.800pt}}
\multiput(481.41,377.00)(0.509,0.533){19}{\rule{0.123pt}{1.062pt}}
\multiput(478.34,377.00)(13.000,11.797){2}{\rule{0.800pt}{0.531pt}}
\multiput(494.41,391.00)(0.509,0.533){19}{\rule{0.123pt}{1.062pt}}
\multiput(491.34,391.00)(13.000,11.797){2}{\rule{0.800pt}{0.531pt}}
\multiput(507.41,405.00)(0.509,0.574){19}{\rule{0.123pt}{1.123pt}}
\multiput(504.34,405.00)(13.000,12.669){2}{\rule{0.800pt}{0.562pt}}
\multiput(520.41,420.00)(0.509,0.574){19}{\rule{0.123pt}{1.123pt}}
\multiput(517.34,420.00)(13.000,12.669){2}{\rule{0.800pt}{0.562pt}}
\multiput(533.41,435.00)(0.509,0.616){19}{\rule{0.123pt}{1.185pt}}
\multiput(530.34,435.00)(13.000,13.541){2}{\rule{0.800pt}{0.592pt}}
\multiput(546.41,451.00)(0.509,0.616){19}{\rule{0.123pt}{1.185pt}}
\multiput(543.34,451.00)(13.000,13.541){2}{\rule{0.800pt}{0.592pt}}
\multiput(559.41,467.00)(0.509,0.657){19}{\rule{0.123pt}{1.246pt}}
\multiput(556.34,467.00)(13.000,14.414){2}{\rule{0.800pt}{0.623pt}}
\multiput(572.41,484.00)(0.509,0.616){19}{\rule{0.123pt}{1.185pt}}
\multiput(569.34,484.00)(13.000,13.541){2}{\rule{0.800pt}{0.592pt}}
\multiput(585.41,500.00)(0.509,0.698){19}{\rule{0.123pt}{1.308pt}}
\multiput(582.34,500.00)(13.000,15.286){2}{\rule{0.800pt}{0.654pt}}
\multiput(598.41,518.00)(0.509,0.657){19}{\rule{0.123pt}{1.246pt}}
\multiput(595.34,518.00)(13.000,14.414){2}{\rule{0.800pt}{0.623pt}}
\multiput(611.41,535.00)(0.509,0.657){19}{\rule{0.123pt}{1.246pt}}
\multiput(608.34,535.00)(13.000,14.414){2}{\rule{0.800pt}{0.623pt}}
\multiput(624.41,552.00)(0.509,0.698){19}{\rule{0.123pt}{1.308pt}}
\multiput(621.34,552.00)(13.000,15.286){2}{\rule{0.800pt}{0.654pt}}
\multiput(637.41,570.00)(0.509,0.698){19}{\rule{0.123pt}{1.308pt}}
\multiput(634.34,570.00)(13.000,15.286){2}{\rule{0.800pt}{0.654pt}}
\multiput(650.41,588.00)(0.509,0.698){19}{\rule{0.123pt}{1.308pt}}
\multiput(647.34,588.00)(13.000,15.286){2}{\rule{0.800pt}{0.654pt}}
\multiput(663.41,606.00)(0.509,0.657){19}{\rule{0.123pt}{1.246pt}}
\multiput(660.34,606.00)(13.000,14.414){2}{\rule{0.800pt}{0.623pt}}
\multiput(676.41,623.00)(0.509,0.657){19}{\rule{0.123pt}{1.246pt}}
\multiput(673.34,623.00)(13.000,14.414){2}{\rule{0.800pt}{0.623pt}}
\multiput(689.41,640.00)(0.509,0.657){19}{\rule{0.123pt}{1.246pt}}
\multiput(686.34,640.00)(13.000,14.414){2}{\rule{0.800pt}{0.623pt}}
\multiput(702.41,657.00)(0.509,0.657){19}{\rule{0.123pt}{1.246pt}}
\multiput(699.34,657.00)(13.000,14.414){2}{\rule{0.800pt}{0.623pt}}
\multiput(715.41,674.00)(0.509,0.616){19}{\rule{0.123pt}{1.185pt}}
\multiput(712.34,674.00)(13.000,13.541){2}{\rule{0.800pt}{0.592pt}}
\multiput(728.41,690.00)(0.509,0.574){19}{\rule{0.123pt}{1.123pt}}
\multiput(725.34,690.00)(13.000,12.669){2}{\rule{0.800pt}{0.562pt}}
\multiput(741.41,705.00)(0.509,0.533){19}{\rule{0.123pt}{1.062pt}}
\multiput(738.34,705.00)(13.000,11.797){2}{\rule{0.800pt}{0.531pt}}
\multiput(754.41,719.00)(0.509,0.533){19}{\rule{0.123pt}{1.062pt}}
\multiput(751.34,719.00)(13.000,11.797){2}{\rule{0.800pt}{0.531pt}}
\multiput(766.00,734.41)(0.536,0.511){17}{\rule{1.067pt}{0.123pt}}
\multiput(766.00,731.34)(10.786,12.000){2}{\rule{0.533pt}{0.800pt}}
\multiput(779.00,746.40)(0.589,0.512){15}{\rule{1.145pt}{0.123pt}}
\multiput(779.00,743.34)(10.623,11.000){2}{\rule{0.573pt}{0.800pt}}
\multiput(792.00,757.40)(0.654,0.514){13}{\rule{1.240pt}{0.124pt}}
\multiput(792.00,754.34)(10.426,10.000){2}{\rule{0.620pt}{0.800pt}}
\multiput(805.00,767.40)(0.737,0.516){11}{\rule{1.356pt}{0.124pt}}
\multiput(805.00,764.34)(10.186,9.000){2}{\rule{0.678pt}{0.800pt}}
\multiput(818.00,776.40)(1.000,0.526){7}{\rule{1.686pt}{0.127pt}}
\multiput(818.00,773.34)(9.501,7.000){2}{\rule{0.843pt}{0.800pt}}
\multiput(831.00,783.38)(1.600,0.560){3}{\rule{2.120pt}{0.135pt}}
\multiput(831.00,780.34)(7.600,5.000){2}{\rule{1.060pt}{0.800pt}}
\multiput(843.00,788.38)(1.768,0.560){3}{\rule{2.280pt}{0.135pt}}
\multiput(843.00,785.34)(8.268,5.000){2}{\rule{1.140pt}{0.800pt}}
\put(856,791.84){\rule{3.132pt}{0.800pt}}
\multiput(856.00,790.34)(6.500,3.000){2}{\rule{1.566pt}{0.800pt}}
\put(869,793.84){\rule{3.132pt}{0.800pt}}
\multiput(869.00,793.34)(6.500,1.000){2}{\rule{1.566pt}{0.800pt}}
\put(882,794.84){\rule{3.132pt}{0.800pt}}
\multiput(882.00,794.34)(6.500,1.000){2}{\rule{1.566pt}{0.800pt}}
\put(895,794.84){\rule{3.132pt}{0.800pt}}
\multiput(895.00,795.34)(6.500,-1.000){2}{\rule{1.566pt}{0.800pt}}
\put(908,793.34){\rule{3.132pt}{0.800pt}}
\multiput(908.00,794.34)(6.500,-2.000){2}{\rule{1.566pt}{0.800pt}}
\put(921,790.84){\rule{3.132pt}{0.800pt}}
\multiput(921.00,792.34)(6.500,-3.000){2}{\rule{1.566pt}{0.800pt}}
\put(934,787.34){\rule{2.800pt}{0.800pt}}
\multiput(934.00,789.34)(7.188,-4.000){2}{\rule{1.400pt}{0.800pt}}
\multiput(947.00,785.06)(1.768,-0.560){3}{\rule{2.280pt}{0.135pt}}
\multiput(947.00,785.34)(8.268,-5.000){2}{\rule{1.140pt}{0.800pt}}
\multiput(960.00,780.06)(1.768,-0.560){3}{\rule{2.280pt}{0.135pt}}
\multiput(960.00,780.34)(8.268,-5.000){2}{\rule{1.140pt}{0.800pt}}
\multiput(973.00,775.07)(1.244,-0.536){5}{\rule{1.933pt}{0.129pt}}
\multiput(973.00,775.34)(8.987,-6.000){2}{\rule{0.967pt}{0.800pt}}
\multiput(986.00,769.07)(1.244,-0.536){5}{\rule{1.933pt}{0.129pt}}
\multiput(986.00,769.34)(8.987,-6.000){2}{\rule{0.967pt}{0.800pt}}
\multiput(999.00,763.08)(1.000,-0.526){7}{\rule{1.686pt}{0.127pt}}
\multiput(999.00,763.34)(9.501,-7.000){2}{\rule{0.843pt}{0.800pt}}
\multiput(1012.00,756.08)(1.000,-0.526){7}{\rule{1.686pt}{0.127pt}}
\multiput(1012.00,756.34)(9.501,-7.000){2}{\rule{0.843pt}{0.800pt}}
\multiput(1025.00,749.07)(1.244,-0.536){5}{\rule{1.933pt}{0.129pt}}
\multiput(1025.00,749.34)(8.987,-6.000){2}{\rule{0.967pt}{0.800pt}}
\multiput(1038.00,743.08)(1.000,-0.526){7}{\rule{1.686pt}{0.127pt}}
\multiput(1038.00,743.34)(9.501,-7.000){2}{\rule{0.843pt}{0.800pt}}
\multiput(1051.00,736.08)(1.000,-0.526){7}{\rule{1.686pt}{0.127pt}}
\multiput(1051.00,736.34)(9.501,-7.000){2}{\rule{0.843pt}{0.800pt}}
\multiput(1064.00,729.07)(1.244,-0.536){5}{\rule{1.933pt}{0.129pt}}
\multiput(1064.00,729.34)(8.987,-6.000){2}{\rule{0.967pt}{0.800pt}}
\multiput(1077.00,723.07)(1.244,-0.536){5}{\rule{1.933pt}{0.129pt}}
\multiput(1077.00,723.34)(8.987,-6.000){2}{\rule{0.967pt}{0.800pt}}
\multiput(1090.00,717.07)(1.244,-0.536){5}{\rule{1.933pt}{0.129pt}}
\multiput(1090.00,717.34)(8.987,-6.000){2}{\rule{0.967pt}{0.800pt}}
\multiput(1103.00,711.07)(1.244,-0.536){5}{\rule{1.933pt}{0.129pt}}
\multiput(1103.00,711.34)(8.987,-6.000){2}{\rule{0.967pt}{0.800pt}}
\multiput(1116.00,705.06)(1.768,-0.560){3}{\rule{2.280pt}{0.135pt}}
\multiput(1116.00,705.34)(8.268,-5.000){2}{\rule{1.140pt}{0.800pt}}
\multiput(1129.00,700.06)(1.768,-0.560){3}{\rule{2.280pt}{0.135pt}}
\multiput(1129.00,700.34)(8.268,-5.000){2}{\rule{1.140pt}{0.800pt}}
\put(1142,693.34){\rule{2.800pt}{0.800pt}}
\multiput(1142.00,695.34)(7.188,-4.000){2}{\rule{1.400pt}{0.800pt}}
\multiput(1155.00,691.06)(1.768,-0.560){3}{\rule{2.280pt}{0.135pt}}
\multiput(1155.00,691.34)(8.268,-5.000){2}{\rule{1.140pt}{0.800pt}}
\put(1168,684.84){\rule{3.132pt}{0.800pt}}
\multiput(1168.00,686.34)(6.500,-3.000){2}{\rule{1.566pt}{0.800pt}}
\put(1181,681.34){\rule{2.800pt}{0.800pt}}
\multiput(1181.00,683.34)(7.188,-4.000){2}{\rule{1.400pt}{0.800pt}}
\put(1194,678.34){\rule{3.132pt}{0.800pt}}
\multiput(1194.00,679.34)(6.500,-2.000){2}{\rule{1.566pt}{0.800pt}}
\put(1207,675.84){\rule{3.132pt}{0.800pt}}
\multiput(1207.00,677.34)(6.500,-3.000){2}{\rule{1.566pt}{0.800pt}}
\put(1220,673.34){\rule{3.132pt}{0.800pt}}
\multiput(1220.00,674.34)(6.500,-2.000){2}{\rule{1.566pt}{0.800pt}}
\put(1233,671.34){\rule{3.132pt}{0.800pt}}
\multiput(1233.00,672.34)(6.500,-2.000){2}{\rule{1.566pt}{0.800pt}}
\put(1246,669.34){\rule{3.132pt}{0.800pt}}
\multiput(1246.00,670.34)(6.500,-2.000){2}{\rule{1.566pt}{0.800pt}}
\put(1259,667.84){\rule{3.132pt}{0.800pt}}
\multiput(1259.00,668.34)(6.500,-1.000){2}{\rule{1.566pt}{0.800pt}}
\put(1272,666.34){\rule{3.132pt}{0.800pt}}
\multiput(1272.00,667.34)(6.500,-2.000){2}{\rule{1.566pt}{0.800pt}}
\put(1285,664.84){\rule{3.132pt}{0.800pt}}
\multiput(1285.00,665.34)(6.500,-1.000){2}{\rule{1.566pt}{0.800pt}}
\put(1311,663.84){\rule{3.132pt}{0.800pt}}
\multiput(1311.00,664.34)(6.500,-1.000){2}{\rule{1.566pt}{0.800pt}}
\put(1298.0,666.0){\rule[-0.400pt]{3.132pt}{0.800pt}}
\put(1337,662.84){\rule{3.132pt}{0.800pt}}
\multiput(1337.00,663.34)(6.500,-1.000){2}{\rule{1.566pt}{0.800pt}}
\put(1324.0,665.0){\rule[-0.400pt]{3.132pt}{0.800pt}}
\put(1402,662.84){\rule{3.132pt}{0.800pt}}
\multiput(1402.00,662.34)(6.500,1.000){2}{\rule{1.566pt}{0.800pt}}
\put(1350.0,664.0){\rule[-0.400pt]{12.527pt}{0.800pt}}
\put(1441,663.84){\rule{3.132pt}{0.800pt}}
\multiput(1441.00,663.34)(6.500,1.000){2}{\rule{1.566pt}{0.800pt}}
\put(1415.0,665.0){\rule[-0.400pt]{6.263pt}{0.800pt}}
\put(1467,664.84){\rule{3.132pt}{0.800pt}}
\multiput(1467.00,664.34)(6.500,1.000){2}{\rule{1.566pt}{0.800pt}}
\put(1454.0,666.0){\rule[-0.400pt]{3.132pt}{0.800pt}}
\end{picture}

\end{center}

\vspace{2cm}

\noindent
{\bf Fig. 1.} 
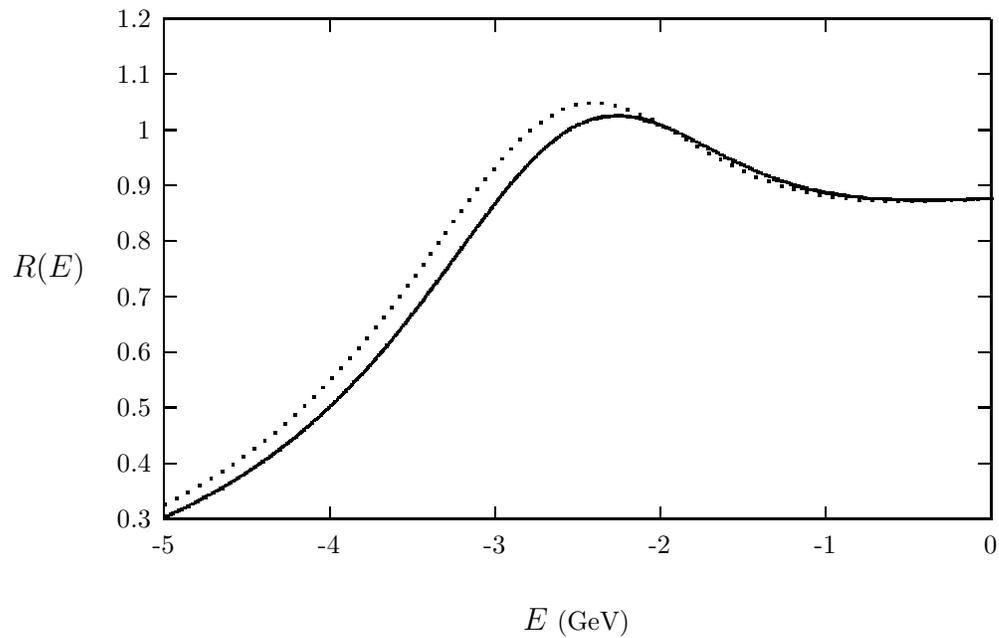
Normalized cross section $R(E)$ of $e^+e^-\to t\bar t$ in NNLO (dotted line)
and with the leading logarithmic N$^3$LO corrections included (solid line), as
a function of the centre-of-mass energy $E$ counted from the nominal threshold
at $2m_t$.

\end{document}